\documentclass[aps,prl,twocolumn,superscriptaddress,nofootinbib,lengthcheck]{revtex4-2}
\usepackage{amsmath,amssymb,amsfonts}
\usepackage{graphicx}
\usepackage{amsmath,amssymb,amsfonts}
\usepackage{graphicx}
\usepackage{dcolumn}
\usepackage{bm}
\usepackage{hyperref}
\usepackage{xcolor}
\usepackage{slashed}
\usepackage{multirow}
\usepackage{booktabs}
\usepackage{tikz-feynhand}   
\usepackage{physics}

\newcommand{\PL}{P_L}
\newcommand{\PR}{P_R}

\newcommand{\vR}{v_R}
\newcommand{\MH}{M_{H^\pm}}
\newcommand{\MN}{M_{N_i}}
\newcommand{\dC}[1]{\Delta C_{#1}^{\mu}}
\newcommand{\Heff}{\mathcal{H}_{\rm eff}}

\newcommand{\Leff}{\mathcal{L}_Y}
\newcommand{\bsll}{b \to s\mu^+\mu^-}
\newcommand{\BKll}{B \to K\mu^+\mu^-}

\begin{document}

\title{Explaining the $B \to K\mu^+\mu^-$ Anomaly in the
       Left-Right Inverse Seesaw Model}

\author{David Delepine}
\email{delepine@ugto.mx}
\affiliation{División de Ciencias e Ingenierías,
  Universidad de Guanajuato, C.P.~37150, León, Guanajuato, México.}

\author{Shaaban Khalil}
\email{skhalil@zewailcity.edu.eg}
\affiliation{Centre for Theoretical Physics,
  Zewail City of Science and Technology,
  6th October City, 12588, Giza, Egypt.}

\date{\today}

\begin{abstract}
We investigate the long-standing anomaly in the rare decay
$\BKll$ within the Left-Right Inverse Seesaw (LRIS) model.
Global analyses of the $\bsll$ data consistently indicate a
significant negative shift in the vector Wilson coefficient,
$\dC{9} \approx -1$, while the axial coefficient $\dC{10}$
remains consistent with zero.
We show that a charged-scalar/heavy-neutrino box diagram in the
LRIS model naturally generates this pattern through a
\emph{non-decoupling} mechanism: the right-handed coupling
 produces a contribution to $\dC{9}$ that is unsuppressed in the heavy-neutrino limit, while the simultaneous presence of a comparable left-handed Dirac Yukawa coupling ensures the automatic cancellation $\dC{10} \approx 0$.
The otherwise large contribution to $B_s$--$\bar{B}_s$ mixing is suppressed by several orders of magnitude through a GIM-like phase structure in the right-handed quark mixing matrix. A numerical scan over the model parameter space identifies a
viable region, consistent with all current flavor and collider constraints.
The $b \to s\gamma$ constraint is satisfied with two orders
of magnitude to spare throughout the viable band.
These results motivate correlated searches for the charged
scalar  and the heavy right-handed neutrinos  at the LHC and future high-luminosity experiments.
\end{abstract}


\maketitle


Flavor-changing neutral-current (FCNC) processes in the
$b \to s$ sector provide among the most sensitive probes
of physics beyond the Standard Model (BSM).
In particular, the rare semileptonic decay $\BKll$ is
forbidden at tree level in the Standard Model (SM) and
proceeds only through suppressed electroweak penguin and
box diagrams~\cite{Glashow:1970gm}.
Its branching fraction and angular observables are therefore
exceptionally sensitive to new particles and interactions
at the TeV scale.

Over the past decade, the LHCb collaboration has reported
a series of measurements revealing systematic deviations
in the angular distributions of $\bsll$ decays compared to SM
predictions~\cite{LHCb:2020lmf,LHCb:2014vgu,LHCb:2022qnv}.
Most notably, the optimised observable $P_5'$ in
$B\to K^*\mu^+\mu^-$ shows persistent discrepancies
at the level of several standard deviations across multiple
$q^2$ bins.
Global analyses of these data consistently indicate a
significant negative shift in the vector Wilson
coefficient~\cite{Altmannshofer:2021qrr,Hurth:2021nsi,
Ciuchini:2022wbq},
\begin{equation}
  \dC{9} \approx -1\,,
  \label{eq:anomaly}
\end{equation}
while the axial coefficient $C_{10}$ remains broadly
consistent with its SM value, $\dC{10} \approx 0$.
This pattern requires a BSM contribution that is predominantly
\emph{vector-like} in the muon current, a feature that is
non-trivial to achieve in concrete models.

In this work, we address this anomaly in the context of
the Left-Right Inverse Seesaw (LRIS)
model~\cite{Pati:1974yy,Mohapatra:1974hk,Senjanovic:1975rk,
Dev:2012sg}, which extends the SM gauge symmetry to
$SU(3)_C \times SU(2)_L \times SU(2)_R \times U(1)_{B-L}$
and incorporates the inverse seesaw mechanism for neutrino
mass generation~\cite{Mohapatra:1986bd,Wyler:1983dd,Pilaftsis:1991ug}.
While the lepton-flavour-universality ratios $R_K$ and
$R_{K^*}$ have recently been found consistent with the
SM~\cite{LHCb:2021trn,LHCb:2022qnv}, the angular
anomalies in $\bsll$ transitions persist and continue to
motivate BSM explanations.
The LRIS model has been studied extensively in the context
of CP violation~\cite{Delepine:2021xxx} and offers a
natural framework for TeV-scale new physics with an
extended scalar sector.

The key mechanism we exploit is a charged-scalar/heavy-neutrino
box diagram that exhibits \emph{non-decoupling} behavior:
the dynamically generated right-handed Yukawa coupling
$Y^R_{\mu N} \propto \MN/\vR$ ensures that the heavy
neutrino mass cancels in the amplitude, leaving a finite,
potentially observable contribution to $\dC{9}$.
Crucially, the simultaneous presence of the left-handed
Dirac Yukawa coupling $Y^L_{\mu N} \sim y^L_{2i}$ with
$|Y^L_{\mu N}| \approx |Y^R_{\mu N}|$ generates equal
and opposite contributions to $C_{10}$, driving $\dC{10}
\approx 0$ automatically — without any additional tuning
beyond the condition natural to the Inverse Seesaw.

The $\bsll$ transition is described by the Weak Effective
Theory (WET) Hamiltonian~\cite{Buchalla:1995vs}
\begin{equation}
  \Heff = -\frac{4G_F}{\sqrt{2}}\,V_{tb}V_{ts}^*\,
  \frac{e^2}{16\pi^2}
  \sum_i \left(C_i\,\mathcal{O}_i
              + C'_i\,\mathcal{O}'_i\right)
  + \text{h.c.}\,,
  \label{eq:Heff}
\end{equation}
where the $\mathcal{O}'_i$ are the chirality-flipped
counterparts of $\mathcal{O}_i$, obtained by $\PL
\leftrightarrow \PR$.
The dominant semileptonic operators are
\begin{align}
  \mathcal{O}_9  &= (\bar{s}\gamma^\mu \PL b)
                    (\bar{\mu}\gamma_\mu\mu)\,,
  \label{eq:O9} \\
  \mathcal{O}_{10} &= (\bar{s}\gamma^\mu \PL b)
                      (\bar{\mu}\gamma_\mu\gamma_5\mu)\,.
  \label{eq:O10}
\end{align}

Global fits to the $\bsll$ data favor the
pattern~\cite{Altmannshofer:2021qrr}
\begin{equation}
  \dC{9} \approx -1\,, \qquad
  \dC{10} \approx 0\,.
  \label{eq:target}
\end{equation}

The chiral origin of this pattern can be understood
model-independently. A purely right-handed muon current
implies $\dC{9} = +\dC{10}$, whereas a purely left-handed
current gives $\dC{9} = -\dC{10}$.
Consequently, neither chiral structure alone can naturally
reproduce the pattern in Eq.~\eqref{eq:target}.
Instead, the preferred scenario requires an approximately
\emph{vector-like} muon current with comparable left-
and right-handed couplings, $g_L^\mu \simeq g_R^\mu$.
In this case,
\begin{align}
  \dC{9}  &\propto |g_L^\mu|^2 + |g_R^\mu|^2 \neq 0\,,
  \label{eq:C9gen} \\
  \dC{10} &\propto |g_R^\mu|^2 - |g_L^\mu|^2 \approx 0\,,
  \label{eq:C10gen}
\end{align}
in the limit $|g_L^\mu| \approx |g_R^\mu|$.

In the SM, $\bsll$ receives contributions from photon
penguin, $Z$-penguin, and $W$-box diagrams, all involving
internal $(t, W^\pm)$ states.
The $W$-box involves only left-handed leptons,
$W^\mu\,\bar{\nu}\gamma_\mu\PL\mu$, generating
\begin{equation}
  \Delta C_9^{\rm box}
  = -\Delta C_{10}^{\rm box}\,,
\end{equation}
while the photon penguin contributes mainly to $C_9$ through
the vector coupling of the photon.
Crucially, the SM does not generate a right-handed muon
current $(\bar{s}\gamma^\mu\PL b)(\bar{\mu}\gamma_\mu\PR\mu)$,
making such a structure a hallmark of BSM physics.


The LRIS model~\cite{Pati:1974yy,Mohapatra:1974hk,
Senjanovic:1975rk,Dev:2012sg} is based on the gauge symmetry
$SU(3)_C \times SU(2)_L \times SU(2)_R \times U(1)_{B-L}$,
with a scalar sector containing a bi-doublet $\phi$ and a
right-handed doublet $\chi_R$.
The symmetry breaking at the scale
$\vR \sim \mathcal{O}(\text{TeV})$
generates the heavy right-handed gauge bosons and heavy
neutrinos, while the bi-doublet vacuum expectation values
generate the electroweak scale and fermion masses.

The Yukawa interactions relevant for our analysis are
\begin{equation}
  \Leff \supset
  y^L_{ij}\,\bar{L}_{Li}\,\phi\,L_{Rj}
  + y^Q_{ij}\,\bar{Q}_{Li}\,\phi\,Q_{Rj}
  + y^S_{ij}\,\bar{L}_{Ri}\,\tilde{\chi}_R\,S^c_{j}
  + \text{h.c.},
  \label{eq:Yukawa}
\end{equation}
where $S_{j}$ are gauge-singlet fermions responsible for
the inverse seesaw mechanism.
This framework naturally accommodates TeV-scale heavy
pseudo-Dirac neutrinos with $\mathcal{O}(1)$ Yukawa
couplings while keeping the light-neutrino masses small
through the inverse-seesaw suppression mechanism~\cite{Dev:2012sg}.
As a result, the heavy neutrinos can have sizable
interactions relevant for flavor observables.

In the scalar sector, two physical charged Higgs bosons
$H^\pm$ remain in the spectrum.
Their interactions with quarks and leptons are determined
by the Yukawa sector in Eq.~\eqref{eq:Yukawa}.
The effective charged-Higgs interactions in the mass basis
can be parameterized as
\begin{eqnarray}
  \mathcal{L} &\supset&
  Y^H_{ts}\,H^+\,\bar{t}_R s_L
  + Y^H_{tb}\,H^+\,\bar{t}_R b_L
  + Y^R_{\mu N_i}\,H^-\,\bar{\mu}_R N_i \nonumber\\
  &+&
  Y^L_{\mu N_i}\,H^-\,\bar{\mu}_L N_i
  + \text{h.c.}\,,
  \label{eq:couplings}
\end{eqnarray}
where $N_i$ denote the heavy pseudo-Dirac neutrinos of
the inverse seesaw sector.
The simultaneous presence of effective left-handed and
right-handed leptonic couplings,
$Y^L_{\mu N_i}$ and $Y^R_{\mu N_i}$,
plays a central role in generating an approximately
vector-like muon current, which is required to enhance
$\dC{9}$ while suppressing $\dC{10}$.

The right-handed coupling originates from the scalar sector
responsible for the breaking of the $SU(2)_R$ gauge symmetry.
In particular, the charged scalar component of the
right-handed multiplet, $\chi_R^\pm$, couples to the heavy
neutrinos through the Yukawa interaction generating their
masses,
\begin{equation}
  M_N \sim y^S \vR.
\end{equation}
After symmetry breaking and rotation to the physical scalar
basis, the charged Higgs boson $H^\pm$ contains a component
of $\chi_R^\pm$,
\begin{equation}
  H^\pm \supset c_R\,\chi_R^\pm,
\end{equation}
where $c_R$ denotes the corresponding mixing coefficient.
As a result, the effective coupling of $H^\pm$ to
$\mu_R$ and $N_i$ becomes proportional to the heavy-neutrino
mass,
\begin{equation}
  Y^R_{\mu N_i}
  \simeq
  c_R\,
  \frac{g_R}{\sqrt{2}}\,
  \frac{\MN}{\vR}\,,
  \label{eq:YR}
\end{equation}
which is analogous to the Goldstone--fermion coupling in
spontaneously broken gauge theories.
Consequently, for heavy neutrino masses of order $\vR$,
the coupling $Y^R_{\mu N_i}$ can naturally be of
$\mathcal{O}(1)$, depending on the charged-scalar mixing
factor $c_R$.

On the other hand, the effective left-handed coupling
originates from the Dirac Yukawa sector and neutrino
mixing effects, with
\begin{equation}
  Y^L_{\mu N_i}
  \sim
  y^L_{2i}\,U_{\nu N}\,,
\end{equation}
where $U_{\nu N}$ denotes the active-heavy neutrino mixing.
In general, the left-handed coupling is model dependent and
can arise after diagonalizing the full neutral-fermion and
charged-scalar mass matrices.


The dominant new-physics contribution to $\bsll$ in the
LRIS framework arises from charged-scalar/heavy-neutrino
box diagrams involving internal $(t,N_i,H^\pm)$ states.
The corresponding amplitudes are generated by the effective
left-handed and right-handed couplings in
Eq.~\eqref{eq:couplings}. After evaluating the loop
integrals, the box amplitude can be written as
{\small
\begin{eqnarray}
  \mathcal{M}_{\rm box}\!\!&=&\!\!
  \frac{1}{16\pi^2}
  \frac{Y^H_{ts}(Y^H_{tb})^*}{\MH^2}
  \Big[
  |Y^R_{\mu N}|^2\,F(x_t,x_N)
  (\bar{s}\gamma^\mu\PL b)
  (\bar{\mu}\gamma_\mu\PR\mu)
  \nonumber\\
  &&\qquad\qquad
  +
  |Y^L_{\mu N}|^2\,F(x_t,x_N)
  (\bar{s}\gamma^\mu\PL b)
  (\bar{\mu}\gamma_\mu\PL\mu)
  \Big],
  \label{eq:Mbox}
\end{eqnarray}}
where $x_t = m_t^2/\MH^2$ and $x_N = \MN^2/\MH^2$,
and $F(x_t,x_N)$ is the scalar box loop function~\cite{Inami:1980fz}.

A remarkable feature of the LRIS contribution is that the
left-handed and right-handed topologies generate the same
loop function $F(x_t,x_N)$.
In the heavy-neutrino limit, the amplitude exhibits a
non-decoupling behavior analogous to the top-quark
contribution in the SM electroweak penguins.
This occurs because the heavy-neutrino dependence in the
loop integral is compensated by the non-decoupling coupling
$Y^R_{\mu N}\propto \MN/\vR$, leaving a finite contribution
suppressed only by the right-handed scale $\vR$.
Consequently, TeV-scale heavy neutrinos can induce sizable
corrections to the Wilson coefficients relevant for
$\bsll$ transitions.


Matching the box amplitude onto the WET Hamiltonian
in Eq.~\eqref{eq:Heff} yields
\begin{align}
  \dC{9}
  &=
  -\frac{v^2\,\mathcal{H}(x_t)}
        {64\pi\alpha_{\rm em}\,\MH^2}
  \frac{Y^H_{ts}(Y^H_{tb})^*}{|V_{ts}|}
  \sum_i
  \left(
    |Y^R_{\mu N_i}|^2
    +
    |Y^L_{\mu N_i}|^2
  \right),
  \label{eq:C9match}
  \\
  \dC{10}
  &=
  -\frac{v^2\,\mathcal{H}(x_t)}
        {64\pi\alpha_{\rm em}\,\MH^2}
  \frac{Y^H_{ts}(Y^H_{tb})^*}{|V_{ts}|}
  \sum_i
  \left(
    |Y^R_{\mu N_i}|^2
    -
    |Y^L_{\mu N_i}|^2
  \right),
  \label{eq:C10match}
\end{align}
where $\mathcal{H}(x_t) \equiv \lim_{x_N\to\infty} x_N\,F(x_t,x_N)$
denotes the finite residue of the loop function in the
heavy-neutrino non-decoupling limit.
Therefore, for approximately equal effective couplings,
\begin{equation}
  |Y^R_{\mu N_i}|
  \simeq
  |Y^L_{\mu N_i}|,
  \label{eq:cancellation}
\end{equation}
the contribution to $\dC{10}$ is naturally suppressed,
while a sizable $\dC{9}$ remains.
Since
\begin{equation}
  Y^R_{\mu N_i}
  \sim
  c_R\,\frac{g_R}{\sqrt2}\,
  \frac{\MN}{\vR},
\end{equation}
with $\MN \sim \vR$, this condition can be naturally
realized for perturbative Dirac Yukawa couplings
$y^L_{2i} = \mathcal{O}(1)$, as expected in the inverse
seesaw framework.

The same couplings driving $\dC{9}$ also contribute to the
$B_s$ mass difference through charged-scalar box diagrams~\cite{Inami:1980fz}.
The experimental value $\Delta M_{B_s}^{\rm exp} = 17.765 \pm 0.006~\text{ps}^{-1}$~\cite{PDG:2022}
sets a stringent constraint on any new contribution.
In the absence of additional flavor structure, the top-quark
contribution would impose strong constraints on
$Y^H_{ts}$ and $Y^H_{tb}$.
However, a GIM-like phase texture in the right-handed quark
mixing sector allows destructive interference between the
top and charm contributions, suppressing
$\Delta M_{B_s}^{\rm NP}$ by several orders of magnitude.
This cancellation does not extend to the semi-leptonic box
amplitude because of the different kinematic structure
induced by the heavy neutrino propagator.

The charged scalar also contributes to the magnetic penguin
operator $\mathcal{O}_7$~\cite{Buras:1993xp}.
For the parameter region relevant to the $\bsll$ anomaly,
the resulting correction remains well below the current
experimental bound on $b\to s\gamma$~\cite{BaBar:2012eja,Belle:2012urf,HFLAV:2022pwe}.

In our numerical analysis, we further impose perturbativity,
requiring the effective Yukawa couplings to remain below
$\mathcal{O}(1)$, together with the LHC direct-search bound
$\MH \gtrsim 600~\text{GeV}$ from
$pp\to H^\pm\to tb$ searches~\cite{CMS:2018dzl,ATLAS:2018ntn},
as well as the lower bound on the right-handed gauge boson mass
$M_{W_R} \gtrsim 4.7~\text{TeV}$~\cite{CMS:2021dzb,ATLAS:2023cjo},
which constrains $\vR$ through $M_{W_R} \simeq g_R\vR/\sqrt{2}$.
The electroweak precision bound $\vR \gtrsim 3~\text{TeV}$ is
also satisfied throughout the viable region~\cite{PDG:2022}.



We perform a numerical scan over the relevant LRIS
parameter space, varying the effective quark couplings
and phases, while fixing the leptonic couplings to
$|Y^L_{\mu N}| = |Y^R_{\mu N}| = 0.8$ in order to realize
the suppression of $\dC{10}$.
The loop amplitudes are evaluated using exact
Passarino--Veltman functions~\cite{Passarino:1978jh}
with $\MH = 1~\text{TeV}$.

We require the parameter points to satisfy the current
$\bsll$ preferred range,
$|\dC{10}| \lesssim 0.2$,
together with the constraints from
$\Delta M_{B_s}$ and $b\to s\gamma$.
A viable region of parameter space is found in which
\begin{equation}
  \dC{9}\sim -1,
  \qquad
  \dC{10}\simeq 0,
\end{equation}
while all flavor constraints remain satisfied.
The allowed solutions are characterized by a GIM-like
phase texture in the right-handed quark sector,
which suppresses the new contribution to
$\Delta M_{B_s}$ through destructive interference between
the top- and charm-quark box amplitudes.


\begin{figure*}[t]
  \centering
  \includegraphics[width=0.8\textwidth]{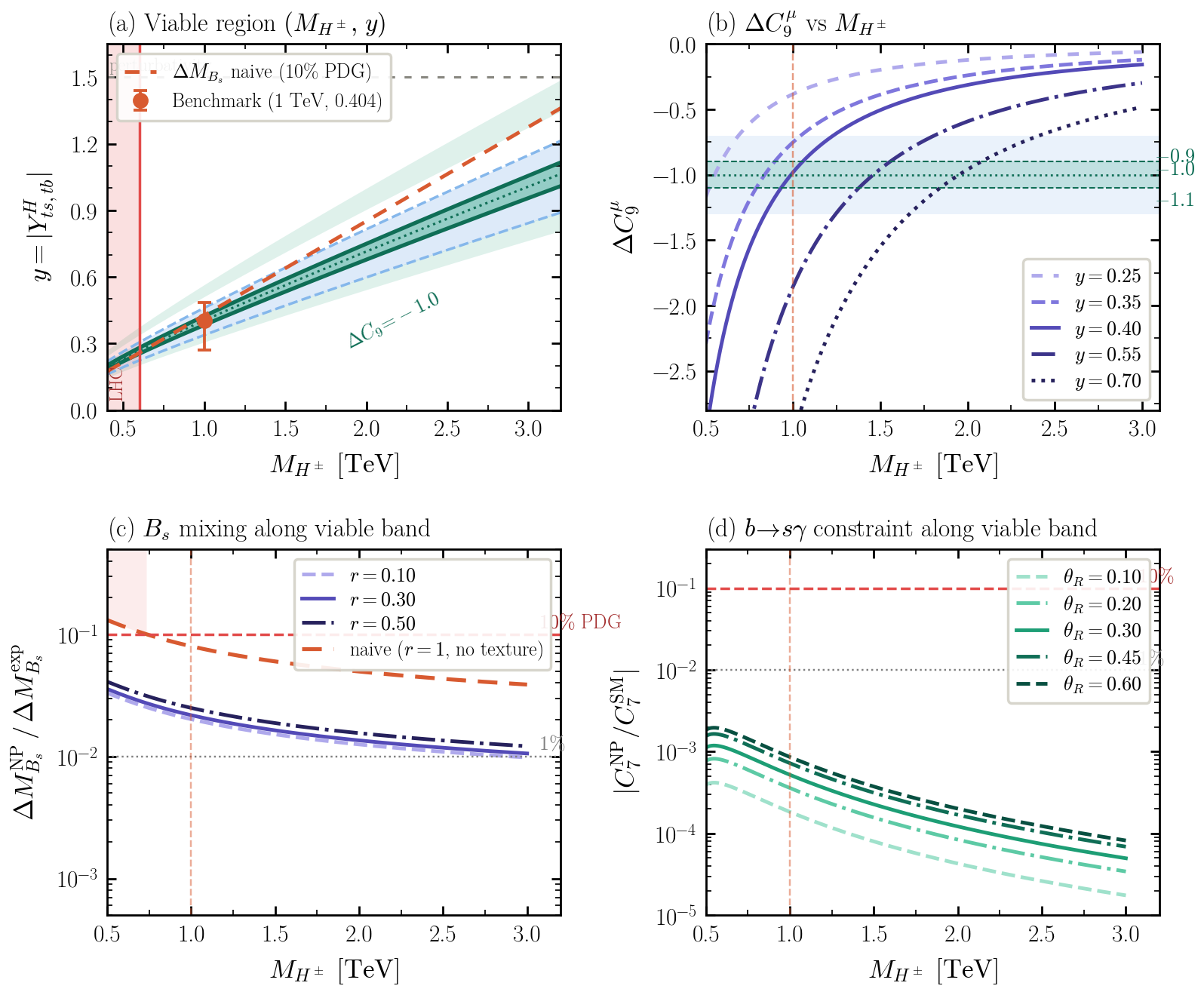}
\caption{
Summary of the LRIS phenomenology and constraints.
(a) Allowed region in the $(M_{H^\pm},y)$ plane.
The dark green (light blue) band corresponds to the
$1\sigma$ ($2\sigma$) preferred range for $\dC{9}$,
while the shaded uncertainty reflects
$y_\mu \in [0.60,1.00]$.
The orange dashed curve indicates the naive
$\Delta M_{B_s}$ constraint in the absence of the
GIM-like phase texture, whereas the red region is excluded
by direct LHC searches for charged Higgs bosons.
(b) $\dC{9}$ as a function of $\MH$ for representative
values of the effective Yukawa coupling $y$.
(c) $\Delta M_{B_s}^{\rm NP}/\Delta M_{B_s}^{\rm exp}$
for the naive (no texture) and phase-texture scenarios,
showing the strong suppression achieved by the GIM-like
cancellation.
(d) The ratio $|C_7^{\rm NP}/C_7^{\rm SM}|$ for different
right-handed mixing angles $\theta_R$.
}
\label{fig:4panel}
\end{figure*}
Figure~\ref{fig:4panel} summarizes the allowed LRIS
parameter space.
Panel~(a) shows the viable region in the
$(M_{H^\pm},y)$ plane, following the expected scaling
$\dC{9}\propto y^2/\MH^2$.
Perturbative solutions remain viable up to
$\MH \sim \mathcal{O}(\text{TeV})$.
Panel~(c) illustrates the strong suppression of
$\Delta M_{B_s}^{\rm NP}$ induced by the GIM-like phase
texture, while panel~(d) shows that the
$b\to s\gamma$ constraint remains negligible throughout
the relevant parameter space.

As a representative benchmark, we consider
\begin{equation}
  \MH = 1~\text{TeV}, \quad
  y = 0.386, \quad
  y_\mu = 0.80, \quad
  \vR = 3~\text{TeV},
\end{equation}
with a CP-violating phase texture in the right-handed
quark sector chosen to maximally suppress $\Delta M_{B_s}$.
This benchmark yields
\begin{equation}
  \dC{9}\simeq -0.91,
  \qquad
  \dC{10}\simeq 0,
\end{equation}
while satisfying the constraints from
$\Delta M_{B_s}$ and $b\to s\gamma$.
The predicted observables are summarized in
Table~\ref{tab:benchmark}.
\begin{table}[h]
\centering
\caption{Benchmark scenario: LRIS model parameters and
predicted observables at $\MH = 1\,\text{TeV}$.
\label{tab:benchmark}}
\begin{ruledtabular}
\begin{tabular}{lll}
\textbf{Parameter / Observable} & \textbf{Value}
& \textbf{Constraint} \\
\colrule
\multicolumn{3}{l}{\textit{Input parameters}} \\
$\MH$                    & $1.0\,\text{TeV}$  & LHC: $>600\,\text{GeV}$ \\
$\vR$                    & $3.0\,\text{TeV}$  & EW precision \\
$y = |Y^H_{ts,tb}|$      & $0.386$            & perturbativity \\
$y_\mu = Y^L = Y^R$      & $0.80$             & non-unitarity \\
CP phase                 & max.\ suppression  & $\Delta M_{B_s} < 10\%$ \\
\colrule
\multicolumn{3}{l}{\textit{Predicted observables}} \\
$\dC{9}$                 & $-0.912$   & $[-1.1,\,-0.9]$~\checkmark \\
$\dC{10}$                & $\approx 0$& $< 0.2$~\checkmark \\
$\Delta M_{B_s}^{\rm NP}/\Delta M_{B_s}^{\rm exp}$
                         & $2.2\%$    & $< 10\%$~\checkmark \\
$|C_7^{\rm NP}/C_7^{\rm SM}|$
                         & $< 10^{-3}$& $< 10\%$~\checkmark \\
\end{tabular}
\end{ruledtabular}
\end{table}

In conclusion, we have shown that the Left-Right Inverse
Seesaw (LRIS) framework can naturally accommodate the
observed $\BKll$ anomaly. The dominant contribution arises from charged-scalar/heavy-
neutrino box diagrams and exhibits several distinctive
features.
First, the non-decoupling coupling
$Y^R_{\mu N_i}\propto \MN/\vR$ generates a sizable shift
in $\dC{9}$ even in the heavy-neutrino limit.
Second, because the left-handed and right-handed box
topologies produce the same loop function, the condition
$|Y^L_{\mu N_i}|\simeq |Y^R_{\mu N_i}|$ naturally leads to
$\dC{10}\simeq 0$ while maintaining
$\dC{9}\sim -1$.
Third, a GIM-like phase texture in the right-handed quark
sector strongly suppresses the contribution to
$\Delta M_{B_s}$ without affecting the semi-leptonic
signal.
Finally, the induced correction to $b\to s\gamma$ remains
well below the current experimental bound throughout the
allowed parameter space.

The viable region extends to charged-Higgs masses in the
multi-TeV range with perturbative Yukawa couplings,
making the scenario testable at present and future LHC
searches through signatures such as
$pp\to H^\pm\to tb$ and
$pp\to W_R\to \ell N$.
Interestingly, the same non-decoupling mechanism also
appears in the LRIS explanation of the forward-backward
CP asymmetry in $\tau\to K\pi\nu_\tau$~\cite{Delepine:2021xxx},
suggesting a common origin for different flavor anomalies
within a unified framework.

\begin{acknowledgments}
The work of S.\,K.\ is partially supported by the Science,
Technology \& Innovation Funding Authority (STDF) under
grant number~48173.
The work of D.\,D.\ is supported by Secretaría de Ciencia,
Humanidades, Tecnología e Innovación (SECIHTI) and
Sistema Nacional de Investigadoras e Investigadores
(S.N.I.I.), Mexico.
\end{acknowledgments}

\end{document}